\documentclass{ifacconf}

\usepackage{graphicx}      % include this line if your document contains figures
\usepackage{natbib}        % required for bibliography

\usepackage{graphics} % for pdf, bitmapped graphics files
\usepackage{epsfig} % for postscript graphics files
\usepackage{algorithm,algorithmic}

\usepackage{balance}
\usepackage{amsmath,amsfonts,amssymb,amscd}
\usepackage{capt-of}
\usepackage{bbm}

\usepackage{bm}

\usepackage{color}
\usepackage{verbatim}
\usepackage{mathtools}

\newcommand{\AR}[1]{\textcolor{blue}{{#1}}}

\newtheorem{remark}{\bfseries Remark}
\newtheorem{theorem}{\bfseries Theorem}
\newtheorem{lemma}{\bfseries Lemma}

\newenvironment{varalgorithm}[1]
  {\algorithm\renewcommand{\thealgorithm}{#1}}
  {\endalgorithm}

\newenvironment{list4}{
	\begin{list}{$\bullet$}{%
			\setlength{\itemsep}{0.05cm}
			\setlength{\labelsep}{0.2cm}
			\setlength{\labelwidth}{0.3cm}
			\setlength{\parsep}{0in} 
			\setlength{\parskip}{0in}
			\setlength{\topsep}{0in} 
			\setlength{\partopsep}{0in}
			\setlength{\leftmargin}{0.16in}}}
	{\end{list}}

\newenvironment{list4a}{
	\begin{list}{$\bullet$}{%
			\setlength{\itemsep}{0.05cm}
			\setlength{\labelsep}{0.2cm}
			\setlength{\labelwidth}{0.3cm}
			\setlength{\parsep}{0in} 
			\setlength{\parskip}{0in}
			\setlength{\topsep}{0in} 
			\setlength{\partopsep}{0in}
			\setlength{\leftmargin}{0.16in}}}
	{\end{list}}

\begin{document}
\begin{frontmatter}

\title{Distributed Optimization with Gradient Descent and Quantized Communication\thanksref{footnoteinfo}} 

\thanks[footnoteinfo]{This work was supported by the Knut and Alice Wallenberg Foundation and the Swedish Research Council. It was also partly supported by the project MINERVA, funded by the European Research Council (ERC) under the European Union's Horizon 2022 research and innovation programme (Grant agreement No. 101044629). \AR{We correct one knowledge point in the blue color for our paper: Distributed optimization with gradient descent and quantized communication. IFAC-PapersOnLine, 56(2), 5900-5906, 2023.}
}

\author[First]{Apostolos~I.~Rikos}  
\author[Second]{Wei~Jiang}
\author[Third]{Themistoklis~Charalambous}
\author[First]{Karl~H.~Johansson}

\address[First]{Division of Decision and Control Systems, KTH Royal Institute of Technology, SE-100 44 Stockholm, Sweden, and Digital Futures, SE-100 44 Stockholm, Sweden. E-mails: {\tt \{rikos,kallej\}@kth.se}}
\address[Second]{Department of Electrical Engineering and Automation, School of Electrical Engineering, Aalto University, Espoo, Finland. E-mail: {\tt wei.jiang@aalto.fi, wjiang.lab@gmail.com}}
\address[Third]{Department of Electrical and Computer Engineering, School of Engineering, University of Cyprus, Nicosia, Cyprus. E-mail: {\tt charalambous.themistoklis@ucy.ac.cy}}

\begin{abstract}                
% Abstract of not more than 250 words.
In this paper, we consider the unconstrained distributed optimization problem, in which the exchange of information in the network is captured by a directed graph topology, thus, nodes can only communicate with their neighbors. 
Additionally, in our problem, the communication channels among the nodes have limited bandwidth. 
In order to alleviate this limitation, quantized messages should be exchanged among the nodes. 
For solving this distributed optimization problem, we combine a gradient descent method with a distributed quantized consensus algorithm (which requires the nodes to exchange quantized messages and converges in a finite number of steps). 
Specifically, at every optimization step, each node (i) performs a gradient descent step (i.e., subtracts the scaled gradient from its current estimate), and (ii) performs a finite-time calculation of the quantized average of every node’s estimate in the network.
As a consequence, this algorithm approximately mimics the centralized gradient descent algorithm. 
We show that our algorithm asymptotically converges to a neighborhood of the optimal solution with linear convergence rate. 
The performance of the proposed algorithm is demonstrated via simple illustrative examples.
\end{abstract}

\begin{keyword}
Distributed optimization, quantized communication, directed graphs, finite-time consensus.
\end{keyword}

\end{frontmatter}
%===============================================================================

% ===============================================
%
%
% INTRODUCTION
%
%
% ===============================================
\section{Introduction}\label{sec:intro}

% \todo{general intro for distr optimization + why it received attention + connections to machine learning}

The problem of distributed optimization has received extensive attention over the recent years from the control and machine learning communities, due to the wide area of applications in research areas such as resource allocation 
\citep{2021:Rikos_Grammenos_Kalyvianaki_Hadj_Themis_Johan_CPU, 2022:Doostmohammadian_Themis_CPU}, sensor networks \citep{2013:Zhu_Guan_OptimCons}, smart grids \citep{2015:Cady}, and federated learning \citep{2020:Jadbabaie_Federated}. 
The main idea is to optimize a global objective function by utilizing multiple nodes over a distributed network. 
Specifically, each node has access to a local function which is part of the global objective function. 
Each node aims to optimize the global objective function by optimizing its own local objective function and then coordinating with other nodes in the network. 
The distributed optimization problem can be defined formally as following: 
\begin{equation}\label{opt_defn}
    \min_{x\in \mathbb{R}^p} F(x) = \sum_{i=1}^n f_i(x) ,
\end{equation}
where $n$ is the number of nodes in the network, $x \in \mathbb{R}^p$ is the decision variable that all nodes are trying to optimize, $f_i : \mathbb{R}^p \mapsto \mathbb{R}$, is the local cost function for every node $v_i = \{1, 2, ..., n\}$, and $F : \mathbb{R}^p \mapsto \mathbb{R}$, is the global cost function. 
Note that due to the distributed nature of the problem, each node communicates only with its neighbors in the underlying network. 
This means that each node is required to perform only local operations such as sensing, communication, and computation in order to cooperatively solve the problem in \eqref{opt_defn}. 

% \todo{Current approaches are (i) primal , (ii) dual. + what are primal, what are dual + in this paper We focus on primal}
% + Salapaka 2020 + Themis 2022 + intenet + 

For solving the distributed problem in \eqref{opt_defn}, there are two main optimization methods: (i) primal-based, and (ii) dual-based. 
% \AR{In primal-based methods, each node utilizes the gradient of it estimate to steer the estimate towards the optimal solution during every iteration.}
% updates its current estimate of the optimal solution by performing (a) a gradient at the current estimate, (b) scaling of the gradient by a specific step, and (c) subtraction of the obtained scaled gradient from the current estimate. \todo{Wei: not sure (a-c) is correct since there are so many different primal based methods for distributed optimzation}
% In this way, the estimates of the nodes are steered towards the optimal solution during every iteration. 
Examples of algorithms which rely on primal-based optimization methods are gradient/subgradient descent (see, e.g., \cite{2009:Nedic_Optim}). 
% In dual-based methods, each node employs Lagrange multipliers. 
% It calculates the gradient of its local function and the gradients of its constraints. 
% Then, it checks the gradient of the local function to point in the same direction as the gradients of its constraints. \todo{Wei: not sure it is correct, why gradient in dual based methods?}
% In this way, each node is able to solve optimization problems by breaking them into smaller pieces. 
Examples of algorithms that rely on dual-based optimization methods are alternating direction method of multipliers (ADMM) algorithms (see, e.g., \cite{2017:Makhdoumi_Ozdaglar,2020:distADMM,ECC2021:Wei-Themis} and references therein). 
%Note that 
In this paper, we focus on primal-based optimization algorithms. %which have less requirements and a simpler structure compared to dual-based ones. 
% \AR{ASK THEMIS}

% \todo{short literature review on primal + mention that extended literature review will be available on extended version of the paper}

There have been various primal-based approaches for  distributed optimization in the literature. 
In \cite{2009:Nedic_Optim}, the authors present a distributed optimization algorithm known as distributed gradient / sub-gradient descent. 
This work converges to the optimal solution with sub-linear convergence rate due to the diminishing step size. 
Improving convergence rate of \cite{2009:Nedic_Optim} is possible by utilizing a constant step size. 
However, utilizing a constant step size leads to convergence to a neighborhood of the optimal solution. 
In \cite{2012:Chen_Ozdaglar, 2014:Jakovetic_Jose}, the authors presented distributed optimization algorithms which execute multiple average consensus steps between each gradient descent update. 
In this way, they achieved faster convergence rate. 
A major drawback of this technique is that the average consensus steps achieve only asymptotic convergence and impose heavy communication requirements over the nodes and the network.  
\cite{2018:Li_Quannan} presented a distributed optimization algorithm which achieves linear convergence rate. 
In order to achieve linear convergence rate, each node relies on historic gradient information and executes one dynamic average consensus step after each gradient descent update. 
In \cite{2017:Nedic_Shi}, the authors present the DIGing algorithm which achieves geometric convergence rate as long as some constraints over the fixed step size are fulfilled. 
In \cite{2015_Shi_Yin}, the authors present the EXTRA algorithm which uses a gradient difference strategy and achieves a geometric convergence rate.
In \cite{2018:Khan_AB}, the authors present a distributed algorithm that geometrically converges to the global minimizer with a sufficiently small step-size. 
The proposed algorithm is based on an inexact gradient method and a gradient estimation technique. 
In \cite{2021:Nedic_PushPull}, the authors present the Push–Pull algorithm which converges to the optimal solution in a linear fashion. 
The Push–Pull algorithm  pushes the information about the gradients to
its neighbors, and pulls information about the decision variable from them. 
%In \cite{2020:Xin_Khan}, the authors present the AB/Push-Pull algorithm which achieves linear. 
%The AB/Push-Pull algorithm utilizes a gradient tracking strategy which is combined with variance reduction. 

% This work was inspired by the following works.

It is important to note that most algorithms in current literature (and all the aforementioned works) assume the processing and exchange of real-valued messages between the nodes in the network. 
%Specifically, nodes exchange the exact values of their calculated parameters in order to guarantee convergence to the desired solution. 
For large-scale networks with possibly limited bandwidth capacity, the communication overhead during each iteration becomes a major bottleneck. 
Quantization of information is one of the major approaches to overcome this issue. 
The main idea of quantization is that nodes transmit a compressed value (i.e., quantized) of their stored information as they require a few bits for representation compared to the non-compressed ones (i.e., real values) which in theory require infinite number of bits. 
For this reason, communication efficient distributed optimization has received significant attention recently in the control and machine learning communities. 
Specifically, in \cite{2015:Ye_Jones}, the authors present a distributed optimization algorithm with progressive quantization. 
They improve the convergence speed via a warm-starting strategy, and show that there exists a trade-off between accuracy and required number of iterations for convergence. 
In \cite{2019:Koloskova_Jaggi} the authors present a gossip-based stochastic gradient descent algorithm, which utilizes arbitrary compressed messages and exhibits linear convergence. 
In \cite{2019:Basu_Diggavi}, the authors present a distributed optimization algorithm which combines aggressive sparsification with quantization. 
The algorithm keeps track of the difference between the true and compressed gradients, and converges with equal convergence rate as its non-quantized version. 
In \cite{2019:Ivkin_Arora}, the authors introduce SKETCHED-SGD. 
This algorithm performs distributed sub-gradient decent (SGD) by communicating sketches instead of full gradients. 
In \cite{2020:Jadbabaie_Federated}, the authors present a communication-efficient federated learning algorithm, which relies on periodic averaging and quantized message-passing, and achieves near-optimal theoretical guarantees. 
In \cite{2020:Li_Chi}, the authors present a distributed optimization algorithm which employs stochastic variance reduction and achieves linear convergence rate. 
In \cite{Khatana:2020ACC}, the authors present a distributed optimization algorithm, called gradient-consensus.
In this algorithm an \emph{approximate} finite-time consensus protocol is combined with gradient descent. 
In \cite{2022:Jiang_Charalambous} the authors propose a distributed algorithm which combines gradient descent and finite-time \emph{exact} ratio consensus. 
Both \cite{Khatana:2020ACC, 2022:Jiang_Charalambous} achieve linear convergence rate.

%Recent developments in the control and machine learning community, have led to an increasing demand of algorithms which are able to optimize a global objective function in a distributed fashion. 
%This means that scientific interest over the topic of communication efficient distributed optimization is set to increase significantly in the near future. 
We emphasize that most current approaches mainly focused on methods which are mainly quantizing values of an asymptotic coordination algorithm {and are only able to exhibit asymptotic convergence to the consensus value, rendering them inappropriate for use in consensus-based distributed optimization methods.} %\todo{Wei: our paper also exhibit asymptotic convergence to the optimal solution, no improvement for this point.}
Notwithstanding this, the problem of how to design communication-efficient algorithms {suitable} for {consensus-based} distributed optimization still remains largely unexplored. 
% \AR{ASK THEMIS}

\vspace{.2cm}

\noindent
\textbf{Main Contributions.}
It is often assumed that the exchange of information among agents is seamless and the exact value is communicated. 
However, in most cases, the exact value is an irrational number, whose transmission would require an infinite number of bits. 
Hence, if the channel has limited capacity, most of the current distributed approaches are {impractical}. 
Additionally, most quantized {consensus} methods are mainly quantizing values of an asymptotic {consensus} algorithm and, as a result, they do not converge in a finite number of steps. 
In this paper, a quantized consensus approach is combined with a gradient decent algorithm yielding the following appealing characteristics:
\begin{list4}
    \item exchange of quantized information, which complies with channels of limited bandwidth;
    %\item finite time for consensus, which facilitates its use between optimization steps in order to compute an approximate value;
    % \item \todo{the operation of the proposed algorithm relies on a novel finite-time quantized averaging strategy, which adjusts to the required quantization level and exhibits distributed stopping capabilites, performing an equivalent to approximate centralized gradient descent iteration to solve the distributed optimization problem.} 
    \item the operation of the proposed algorithm relies on a novel finite-time quantized averaging strategy. 
    The averaging strategy adjusts to the required quantization level and exhibits distributed stopping capabilites.
    Our algorithm's performance is equivalent to approximate centralized gradient descent iteration to solve the distributed optimization problem.
\end{list4}
The main idea behind our proposed algorithm is the following. 
Initially, each node stores its quantized estimation regarding the optimal solution. 
During the algorithm's operation, each node performs a gradient descent step (i.e., subtracts the scaled gradient from its current estimate). 
Then, each node performs a finite-time calculation of the quantized average of every node's estimate in the network. 
This operation allows our algorithm to asymptotically converge to a neighborhood of the optimal solution with linear convergence rate.
% \todo{Wei: repetitive, not needed} \AR{we have space, we could keep it}

This work was inspired by \cite{Khatana:2020ACC, 2022:Jiang_Charalambous}. 
% as it performs an approximate (exact) centralized \todo{cannot prove it is centralized - \AR{IS MENTIONED IN KHATANA PAPER}} gradient descent to solve the distributed optimization problem. 
However, our algorithm is substantially different than \cite{2022:Jiang_Charalambous, Khatana:2020ACC} in terms of operation and operational advantages. 
Specifically, the operation of our proposed algorithm relies on a novel quantized averaging algorithm adjusted to the desired quantization level. 
Furthermore, due to its quantized operation, in practical scenarios our proposed algorithm may exhibit (i) reduced complexity (i.e., quantized values can be represented using fewer bits than real values), (ii) faster computation (i.e., operating on quantized values can be faster than operating on real values because the former requires fewer computational resources), and (iii) improved accuracy in certain cases such as processing signals that have low signal-to-noise ratios (i.e., in some cases, using quantized values can actually improve the accuracy of the algorithm, e.g., when processing signals that have low signal-to-noise ratios, quantization can help to reduce the effects of noise and improve the signal quality).
%operates with quantized values.
%The main differences are that the proposed algorithm (i) operates with quantized values, and (ii) exhibits finite time convergence. 
%Furthermore, the proposed algorithms’ operation relies on a novel quantized averaging algorithm whose operation is adjusted to the desired quantization level. 

% ===============================================
%
%
% NOTATION
%
%
% ===============================================
\section{NOTATION AND BACKGROUND}\label{sec:preliminaries}

\noindent
\textbf{Notation.}
We denote the following sets of numbers: real $\mathbb{R}$, rational $\mathbb{Q}$, integer $\mathbb{Z}$, and natural $\mathbb{N}$. 
The set of nonnegative integers is denoted as $\mathbb{Z}_{\geq 0}$. 
The set of positive rationals is denoted as $\mathbb{Q}_{> 0}$. 
For any $a \in \mathbb{R}$, the greatest integer less than or equal to $a$ is denoted $\lfloor a \rfloor$, and the least integer greater than or equal to $a$ is denoted as $\lceil a \rceil$. 
Matrices are denoted with capital letters (e.g., $A$), and vectors with small letters (e.g., $x$). 
The transpose of matrix $A$ and vector $x$ are denoted as $A^\top$, $x^\top$, respectively. 
The Euclidean norm of a vector is denoted as $\| x \|$. 
% For a matrix $A\in \mathbb{R}^{n\times n}$, the entry at row $i$ and column $j$ is denoted by $A_{ij}$. 
By $\mathbf{1}$ we denote the all-ones vector and by $I$ we denote the identity matrix (of appropriate dimensions). 
%By $e_j$ we denote a column vector of all $0$ whose $j^{th}$ entry is $1$. 
By $\nabla$ we denote the standard derivative of a function. 

\vspace{.2cm}

\noindent
\textbf{Graph Theory.}
The communication topology of the network consists of $n$ ($n \geq 2$) nodes communicating only with their immediate neighbors. 
This can be captured by a directed graph (digraph) defined as $\mathcal{G} = (\mathcal{V}, \mathcal{E})$. 
In $\mathcal{G}$, the set of nodes is denoted as $\mathcal{V} =  \{v_1, v_2, \dots, v_n\}$, and the set of edges as $\mathcal{E} \subseteq \mathcal{V} \times \mathcal{V} \cup \{ (v_j, v_j) \ | \ v_j \in \mathcal{V} \}$ (note that each node has also a virtual self-edge). 
% In $\mathcal{G}$, each edge $(v_j, v_i) \in \mathcal{E}$ admits a nonnegative weight $w_{ji} \in \mathbb{N}$ (note that if $(v_j, v_i) \notin \mathcal{E}$, then $w_{ji} = 0$, for every $v_j, v_i \in \mathcal{V}$). 
% The set of edge weights is denoted as $\mathcal{W} = \{ w_{ji} \ | \ (v_j, v_i) \in \mathcal{E} \}$. 
The cardinality of the set of nodes is denoted as $| \mathcal{V} |  = n$, and the cardinality of the set of edges as $m = | \mathcal{E} |$. 
A directed edge from node $v_i$ to node $v_j$ is denoted by $m_{ji} \triangleq (v_j, v_i) \in \mathcal{E}$, and captures the fact that node $v_j$ can receive information from node $v_i$ (but not the other way around). 
The subset of nodes that can directly transmit information to node $v_j$ is called the set of in-neighbors of $v_j$ and is represented by $\mathcal{N}_j^- = \{ v_i \in \mathcal{V} \; | \; (v_j,v_i)\in \mathcal{E}\}$. 
The cardinality of $\mathcal{N}_j^-$ is called the \textit{in-degree} of $v_j$ and is denoted by $\mathcal{D}_j^- = | \mathcal{N}_j^- |$. 
The subset of nodes that can directly receive information from node $v_j$ is called the set of out-neighbors of $v_j$ and is represented by $\mathcal{N}_j^+ = \{ v_l \in \mathcal{V} \; | \; (v_l,v_j)\in \mathcal{E}\}$. 
The cardinality of $\mathcal{N}_j^+$ is called the \textit{out-degree} of $v_j$ and is denoted by $\mathcal{D}_j^+ = | \mathcal{N}_j^+ |$. 
A directed \textit{path} of length $t$ from $v_i$ to $v_j$ exists if we can find a sequence of nodes $v_i \equiv v_{l_0},v_{l_1}, \dots, v_{l_t} \equiv v_j$ such that $(v_{l_{\tau+1}},v_{l_{\tau}}) \in \mathcal{E}$ for $ \tau = 0, 1, \dots , t-1$. 
The diameter $D$ of a digraph is the longest shortest path between any two nodes $v_j, v_i \in \mathcal{V}$ in the network. 

\vspace{.2cm}

\noindent
\textbf{Node Operation.}
At each time step $k \in \mathbb{Z}_{\geq 0}$ each node $v_j$ maintains: (i) its local estimate variable $x_j^{[k]} \in \mathbb{Q}$ which is used to calculate the optimal solution, (ii) the stopping variables $M_j$, $m_j \in \mathbb{Q}$, which are used to determine whether convergence has been achieved, (iii) the mass variables $y_j^{[k]} \in \mathbb{Q}$ and $z_j^{[k]} \in \mathbb{Q}$, which are used to communicate with other nodes by either transmitting or receiving messages, and (iv) the state variables $y_{j,(s)}^{[k]} \in \mathbb{Q}$, $z_{j,(s)}^{[k]} \in \mathbb{Q}$ and $q_{j,(s)}^{[k]} = y_{j,(s)}^{[k]} / z_{j,(s)}^{[k]}$, which are used to store the received messages and calculate the result of the optimization operation. 
% the transmission variables $S\_br_j \in \mathbb{N}$ and $M\_tr_j \in \mathbb{N}$, which are used to decide whether $v_j$ will broadcast its state variables or transmit its mass variables via a direct transmission. 

\vspace{.2cm}

\noindent
\textbf{Synchronous $\max$/$\min$ - Consensus.} 
The distributed $\max$-consensus algorithm computes the maximum value of the network in a finite number of time steps (see \cite{2008:Cortes}). 
Every node $v_{j} \in \mathcal{V}$ updates its state in a synchronous fashion with the following update rule: 
\begin{align}\label{max_operation_eq}
q_{j}^{[k+1]} = \max_{v_{i}\in \mathcal{N}_j^{-} \cup \{v_{j}\}}\{ q_{i}^{[k]} \}.
\end{align}
The $\max$-consensus algorithm converges to the maximum value among all nodes in a finite number of steps $s_m$, where $s_m \leq D$  (see, \cite[Theorem 5.4]{2013:Giannini}).  
Similar results hold for the $\min$-consensus algorithm.

\vspace{.2cm}

\noindent
\textbf{Asymmetric Quantizer.}
In distributed networks, quantization is a common procedure to reduce the required communication bandwidth and to increase power and computation efficiency. 
Quantization lessens the number of bits needed to represent information. 
It is mainly used to describe communication constraints and imperfect information exchanges between nodes \citep{2019:Wei_Johansson}. 
The three main types of quantizers are (i) uniform, (ii) asymmetric, and (iii) logarithmic. 
In this paper we rely on asymmetric quantizers to lessen the number of bits needed to represent information (but the results can also be extended to logarithmic and uniform quantizers). 
Assymetric quantizers are defined as 
\begin{equation}\label{asy_quant_defn}
    q_{\Delta}^a(\xi) = \Bigl \lfloor \frac{\xi}{\Delta} \Bigr \rfloor \Delta, 
\end{equation}
where $\xi \in \mathbb{R}$ is the value to be quantized, $q_{\Delta}^a(\xi) \in \mathbb{Q}$ is the quantized version of $\xi$, and $\Delta \in \mathbb{Q}$ is the quantization level.

% ===============================================
%
%
% PROBLEM
% 
%
% ===============================================
\section{Problem Formulation}\label{sec:probForm}

% \todo{we calculate neighborhood. mention this?}

Let us consider a digraph $\mathcal{G} = (\mathcal{V}, \mathcal{E})$ with $n  = | \mathcal{V} |$ nodes. 
Each node $v_j$ is endowed with a local cost function $f_j(x): \mathbb{R}^P \mapsto \mathbb{R}$ only known to node $v_j$. 
We aim to develop a distributed algorithm which allows nodes to cooperatively solve the following optimization problem $P1$: 
\begin{subequations}
\begin{align}
\min_{x\in \mathcal{X}}~ & F(x_1, x_2, ..., x_n) \equiv \sum_{i=1}^n f_i(x_i), \label{Global_cost_function}  \\
\text{s.t.}~ & x_i = x_j, \forall v_i, v_j, \in \mathcal{V}, \label{constr_same_x}  \\
       & x_i^{[0]} \in \mathcal{X} \subset \mathbb{Q}_{\geq 0}, \forall v_i \in \mathcal{V}, \label{constr_x_in_X} \\
       & \text{nodes communicate with quantized values. } \label{constr_quant} 
       % & \text{distributed termination after convergence.} \label{constr_stop}
%\end{split}\tag{P1}
\end{align}
\end{subequations}
We denote $\mathcal{X}$ the set of feasible values of parameter $x$, and $x^*$ the optimal solution of the optimization problem.  
Eq.~\eqref{Global_cost_function} means that we aim to minimize the global cost function which is defined as the sum of the local cost functions in the network. 
Eq.~\eqref{constr_same_x} means that nodes need to calculate equal optimal solutions. 
Eq.~\eqref{constr_x_in_X} means that the initial estimations of nodes belong in a common set. 
Note that it is not necessary for the initial values of nodes to be rational numbers, i.e., $x_i^{[0]} \in \mathcal{X} \subset \mathbb{Q}_{\geq 0}$. 
However, nodes can generate a quantized version of their initial states by utilizing the Asymetric Quantizer presented in Section~\ref{sec:preliminaries}.
Eq.~\eqref{constr_quant} means that nodes are transmitting and receiving quantized values with their neighbors. 
% \todo{This section is too short; whether merge it with Section 4 or do some changes.}
% Eq.~\eqref{constr_stop} means that nodes are able to distributively determine whether convergence has been achieved and terminate their operation. 

% \todo{explain all 3a until 3e}
% In the remainder of the paper, the problem described in \eqref{Global_cost_functions}--\eqref{Global_cost_functions2} is denoted as $P1$. 

% \todo{say that you have 1 dimension but results can be easily extended to more  dimensions}

% ===============================================
%
%
% ALGORITHM
%
%
% ===============================================
\section{Finite Time Distributed Optimization with Averaged Quantized Gradients}\label{sec:Quant_Grad_Alg}

In this section we present a distributed algorithm which solves the problem described in Section~\ref{sec:probForm}. 
Our distributed algorithm is detailed below as Algorithm~\ref{algorithm_1} (with name QuAGD). 
Before presenting QuAGD, we consider the following assumptions for the development of the results in this paper. 

\begin{assum}\label{str_conn}
We assume $\mathcal{G}$ is \textit{strongly connected}. 
This means that there exists a directed \textit{path} from $v_i$ to $v_j$, for every $v_j, v_i \in \{ \mathcal{V} \ \vert \ v_j \neq v_i\}$. 
\end{assum}

\begin{assum}\label{lipsch_str_conv}
    For every node $v_j$, the local cost function $f_j(x)$ is smooth and strongly convex. 
    This means that for every node $v_j$, for every $x_1, x_2, \in \mathcal{X}$, 
    \begin{itemize}
        \item there exists positive constant $L_j$ such that
        \begin{equation}\label{lipschitz_defn}
            \| \nabla f_j(x_1) - \nabla f_j(x_2) \|_2 \leq L_j \| x_1 - x_2 \|_2, 
        \end{equation}
        \item there exists positive constant $\mu_j$ such that 
        \begin{equation}\label{str_conv_defn}
             f_j(x_2) \geq f_j(x_1) + \nabla f_j(x_1)^\top (x_2 - x_1) + \frac{\mu_j}{2} \| x_2 - x_1 \|_2^2. 
        \end{equation}
    \end{itemize}
    This means that the Lipschitz-continuity and strong-convexity constants of the global cost function $F$ (see \eqref{Global_cost_function}) are $L$, $\mu$, respectively ($L$, $\mu$ are defined later in Theorem~\ref{theorem_convergence_stronglyConvex}).
\end{assum}

\begin{assum}\label{Diam_known}
Every node $v_j \in \mathcal{V}$ knows the diameter of the network $D$ or an upper bound $D'$ (i.e., $D' \geq D$), and a common quantization level $\Delta$. 
\end{assum}

Assumption~\ref{str_conn} is a necessary condition so that each node is able to calculate the optimal solution $x^*$ of $P1$. 

In assumption~\ref{lipsch_str_conv}, Lipschitz-continuity (see \eqref{lipschitz_defn}) is a necessary condition that guarantees the existence of the solution. 
% It limits the value of the gradient of a function (i.e., how fast a function can change). 
Lipschitz-continuity is a standard assumption in distributed first-order optimization problems (see \cite{2018:Xie, 2018:Li_Quannan}) and guarantees that nodes are able to calculate the global optimal minimizer $x^*$ for \eqref{Global_cost_function}. 
Also, strong-convexity (see \eqref{str_conv_defn}) is useful for guaranteeing a linear convergence rate and that the global function $F$ has no more than one minimum. 
% \todo{confused, I think some parts are not precise.} 

Assumption~\ref{Diam_known} allows each node to determine whether it has calculated the quantized average of every node’s estimate of the optimal solution in finite time (and then proceed to perform gradient descent).  

\subsection{Quantized Averaged Gradient Descent Algorithm}\label{alg_code}

The details of the distributed optimization algorithm can be seen in Algorithm~\ref{algorithm_1}. 

% \todo{write mass summation with smarter distributed stopping which depends on the mass merge and not the node states??}

\noindent
\vspace{-0.5cm}    
\begin{varalgorithm}{1}
\caption{Quantized Averaged Gradient Descent (QuAGD)}
\textbf{Input:} A strongly connected digraph $\mathcal{G}$ with $n = |\mathcal{V}|$ nodes and $m = |\mathcal{E}|$ edges. 
% Row stochastic matrix $\overline{B}^{n \times n} = [\overline{b}_{ji}]^{n \times n}$ ($\overline{b}_{ji}>0$ if $(v_j, v_i) \in \mathcal{E}$, $0$ otherwise). 
Static step-size $\alpha \in \mathbb{R}$, digraph diameter $D$, initial value $x_j^{[0]}$, local cost function $f_j$, quantization level $\Delta \in \mathbb{Q}$, for every node $v_j \in \mathcal{V}$.  
\\
% \textbf{Initialization:} Each node $v_j \in \mathcal{V}$ sets $q_{j,(s)}^{[0]} = \nabla f_j (x_j^{[0]})$, $x_j^{[0]} = \Bigl \lfloor \frac{x_j^{[0]}}{\Delta} \Bigr \rfloor \Delta$.
% \\ 
\textbf{Iteration:} For $k=0,1,2,\dots$, each node $v_j \in \mathcal{V}$ does: 
\begin{list4}
    % \item[$1)$] $x_j^{[k+\frac{1}{2}]} = \sum_{i=1}^n \overline{b}_{ji} x_i^{[k]} - \alpha q_{j,(s)}^{[k]} $;\text{ old}
    % \item[$1.5)$] $x_j^{[k+1]} = \Delta \sum_{i=1}^n \overline{b}_{ji} \Bigl \lfloor \frac{ x_i^{[k]}}{\Delta} \Bigr \rfloor - \alpha q_{j,(s)}^{[k]}$; Consider
    %\item[$2)$] $x_j[k+\frac{2}{3}] = \frac{x_j[k+\frac{1}{3}]}{\Delta} + e_{j}^q$;
    %\item[$3)$] $e_{j}^q = x_j[k+\frac{2}{3}] - \lfloor x_j[k+\frac{2}{3}] \rfloor$; 
    %\item[$4)$] $x_j[k+1] = \lfloor x_j[k+\frac{2}{3}] \rfloor \Delta$; 
    % \item[$2)$] $q_{j,(s)}^{[k+1]} =$ Algorithm~\ref{algorithm_1a}($\Bigl \lfloor \frac{\nabla f_j(x_j^{[k+\frac{1}{2}]})}{\Delta} \Bigr \rfloor, D, \Delta $); \text{ old}
    % \item[$3)$] $x_j^{[k+1]} = \Bigl \lfloor \frac{x_j^{[k+\frac{1}{2}]}}{\Delta} \Bigr \rfloor \Delta $. \text{ old}
    \item[$1)$] $x_j^{[k+\frac{1}{2}]} =  x_j^{[k]} - \alpha \nabla f_j(x_j^{[k]})$;
    \item[$2)$] $x_j^{[k+1]} =$ Algorithm~\ref{algorithm_1a}($x_j^{[k+\frac{1}{2}]}, D, \Delta $); 
\end{list4}
\textbf{Output:} Each node $v_j \in \mathcal{V}$ calculates $x^*$ which solves problem $P1$ in Section~\ref{sec:probForm}. 
\label{algorithm_1}
\end{varalgorithm}

\noindent
\vspace{-0.5cm}    
\begin{varalgorithm}{1a}
\caption{FAQuA}
\textbf{Input:} $x_j^{[k+\frac{1}{2}]}, D, \Delta$. 
\textbf{Output:} $x_j^{[k+1]}$. 
\\
\textbf{Initialization:} Each node $v_j \in \mathcal{V}$ does the following: 
\begin{list4}
\item[$1)$] Assigns a nonzero probability $b_{lj}$ to each of its outgoing edges $m_{lj}$, where $v_l \in \mathcal{N}^+_j \cup \{v_j\}$, as follows
\begin{align*}
b_{lj} = \left\{ \begin{array}{ll}
         \frac{1}{1 + \mathcal{D}_j^+}, & \mbox{if $l = j$ or $v_{l} \in \mathcal{N}_j^+$,}\\
         0, & \mbox{if $l \neq j$ and $v_{l} \notin \mathcal{N}_j^+$,}\end{array} \right. 
\end{align*} 
\item[$2)$] $\text{flag}_j = 0$, $z_j^{[1]} = 2$, $y_j^{[1]} = 2 \Bigl \lfloor \frac{x_j^{[k+\frac{1}{2}]}}{\Delta} \Bigr \rfloor$, 
\item[$3)$] $y_{j,(s)}^{[1]} := y_j^{[1]}$, $z_{j,(s)}^{[1]} = z_j^{[1]}$, $q_{j,(s)}^{[1]} := y_{j,(s)}^{[1]} / z_{j,(s)}^{[1]}$, 
\item[$4)$] chooses $v_l \in \mathcal{N}^+_j \cup \{v_j\}$ randomly according to $b_{lj}$, and transmits $y_j^{[1]}$ and $z_j^{[1]}$ towards $v_l$. 
\end{list4} 
\textbf{Iteration:} For $\lambda = 1,2,\dots$, each node $v_j \in \mathcal{V}$, does: 
% \begin{list4} 
% \item \textbf{while} $\text{flag}_j = 0$ \textbf{then} 
\begin{list4a}
\item[$1)$] \textbf{if} $\lambda \mod D = 1$ \textbf{then} sets $M_j = \lceil y_{j,(s)}^{[\lambda]} / z_{j,(s)}^{[\lambda]} \rceil$, $m_j = \lfloor y_{j,(s)}^{[\lambda]} / z_{j,(s)}^{[\lambda]} \rfloor$; 
\item[$2)$] broadcasts $M_j$, $m_j$ to every $v_{l} \in \mathcal{N}_j^+$; 
\item[$3)$] receives $M_i$, $m_i$ from every $v_{i} \in \mathcal{N}_j^-$; 
\item[$4)$] sets $M_j = \max_{v_{i} \in \mathcal{N}_j^-\cup \{ v_j \}} M_i$, \\ $m_j = \min_{v_{i} \in \mathcal{N}_j^-\cup \{ v_j \}} m_i$; 
\item[$5)$] \textbf{if} $z_{j}^{[\lambda]} > 1$, \textbf{then} 
\begin{list4a}
\item[$5.1)$] sets $z_{j,(s)}^{[\lambda]} = z_{j}^{[\lambda]}$, $y_{j,(s)}^{[\lambda]} = y_{j}^{[\lambda]}$, 
$
q_{j,(s)}^{[\lambda]} = \Bigl \lfloor \frac{y_{j,(s)}^{[\lambda]}}{z_{j,(s)}^{[\lambda]}} \Bigr \rfloor \ ;
$
\item[$5.2)$] sets (i) $mas^{y,[\lambda]} = y_{j}^{[\lambda]}$, $mas^{z,[\lambda]} = z_{j}^{[\lambda]}$; (ii) $c^{y,[\lambda]}_{lj} = 0$, $c^{z,[\lambda]}_{lj} = 0$, for every $v_l \in \mathcal{N}^+_j \cup \{v_j\}$; (iii) $\delta = \lfloor mas^{y[\lambda]} / mas^{z,[\lambda]} \rfloor$, $mas^{rem,[\lambda]}= y_{j}^{[\lambda]} - \delta \ mas^{z,[\lambda]}$;  
\item[$5.3)$] \textbf{while} $mas^{z,[\lambda]} > 1$, \textbf{then} 
\begin{list4a}
\item[$5.3a)$] chooses $v_l \in \mathcal{N}^+_j \cup \{v_j\}$ randomly according to $b_{lj}$; 
\item[$5.3b)$] sets (i) $c^{z,[\lambda]}_{lj} := c^{z,[\lambda]}_{lj} + 1$, $c^{y,[\lambda]}_{lj} := c^{y,[\lambda]}_{lj} + \delta$; (ii) $mas^{z,[\lambda]} := mas^{z,[\lambda]} - 1$, $mas^{y,[\lambda]} := mas^{y,[\lambda]} - \delta$. 
\item[$5.3c)$] If $mas^{rem,[\lambda]} > 1$, sets $c^{y,[\lambda]}_{lj} := c^{y,[\lambda]}_{lj} + 1$, $mas^{rem,[\lambda]} := mas^{rem,[\lambda]} - 1$; 
\end{list4a}
\item[$5.4)$] sets $c^{y,[\lambda]}_{jj} := c^{y,[\lambda]}_{jj} + mas^{y,[\lambda]}$, $c^{z,[\lambda]}_{jj} := c^{z,[\lambda]}_{jj} + mas^{z,[\lambda]}$; 
\item[$5.5)$] for every $v_l \in \mathcal{N}^+_j$, if $c^{z,[\lambda]}_{lj} > 0$ transmits $c^{y,[\lambda]}_{lj}$, $c^{z,[\lambda]}_{lj}$ to out-neighbor $v_l$; 
\end{list4a}
\item \textbf{else if} $z_{j}^{[\lambda]} \leq 1$, sets $c^{y,[\lambda]}_{jj} = y_{j}^{[\lambda]}$, $c^{z,[\lambda]}_{jj} = z_{j}^{[\lambda]}$; 
\item[$6)$] receives $c^{y,[\lambda]}_{ji}$, $c^{z,[\lambda]}_{ji}$ from $v_i \in \mathcal{N}_j^-$ and sets 
\begin{equation}\label{no_del_eq_y_no_oscil}
y_{j}^{[\lambda+1]} = c^{y,[\lambda]}_{jj} + \sum_{i=1}^{n}  w_{ji}^{[\lambda]} \ c^{y,[\lambda]}_{ji} ,
\end{equation} 
\begin{equation}\label{no_del_eq_z_no_oscil}
z_{j}^{[\lambda+1]} = c^{z,[\lambda]}_{jj} + \sum_{i=1}^{n} w_{ji}^{[\lambda]} \ c^{z,[\lambda]}_{ji} ,
\end{equation}
where $w_{ji}^{[\lambda]} = 1$ if node $v_j$ receives $c^{y,[\lambda]}_{jj}$, $c^{z,[\lambda]}_{jj}$ from $v_i \in \mathcal{N}_j^-$ at iteration $k$ (otherwise $w_{ji}^{[\lambda]} = 0$); 
\item[$7)$] \textbf{if} $\lambda \mod D = 0$ \textbf{then}, \textbf{if} $M_j - m_j \leq 1$ \textbf{then} sets $x_j^{[k+1]} = m_j \Delta$ and stops operation. 
\end{list4a}
% \end{list4}
\label{algorithm_1a}
\end{varalgorithm}

The intuition of Algorithm~\ref{algorithm_1} (QuAGD) is the following. 
Initially, each node maintains an estimate of the optimal solution, and the desired quantization level. 
Quantization level (i) is the same for every node, (ii) allows quantized communication between nodes, and (iii) determines the desired precision of the solution. 
At each time step $k$, each node updates the estimate of the optimal solution by performing a gradient descent step. 
This step is performed towards the negative direction the node's gradient. 
Then, each node utilizes a fast asymmetrically quantized averaging algorithm\footnote{Algorithm~\ref{algorithm_1a} (FAQuA) runs between every two consecutive optimization steps $k$ and $k + 1$ of Algorithm~\ref{algorithm_1} (QuAGD). For this reason Algorithm~\ref{algorithm_1a} uses a different time index $\lambda$ (and not $k$ as Algorithm~\ref{algorithm_1}).} Algorithm~\ref{algorithm_1a} (FAQuA).
FAQuA allows each node to update its estimate of the optimal solution. 
Specifically, FAQuA allows each node to calculate the quantized average of each node’s estimate in finite time by processing and transmitting quantized messages, with precision determined by the quantization level.
The intuition of FAQuA is explained below. 

The intuition of Algorithm~\ref{algorithm_1a} (FAQuA) is the following. 
FAQuA algorithm utilizes (i) asymmetric quantization, (ii) quantized averaging, and (iii) a stopping strategy. 
Specifically, each node $v_j$ uses an asymmetric quantizer to its state and doubles its mass variables (this change has no effect on the average calculation). 
Then, at each time step $\lambda$, each node $v_j$ checks if $z_j[\lambda] > 1$ (i) it updates its state variables to be equal to the mass variables and (ii) it \textit{splits} $y_j[\lambda]$ into $z_j[\lambda]$ equal integer pieces (the value of some pieces might be greater than others by one). 
It chooses one piece with minimum $y$-value and transmits it to itself, and it transmits each of the remaining $z_j[\lambda]-1$ pieces to randomly selected out-neighbors or to itself. 
It receives the values $y_i[\lambda]$ and $z_i[\lambda]$ from its in-neighbors, sums them with its stored $y_j[\lambda]$ and $z_j[\lambda]$ values and repeats the operation. 
Every $D$ time steps performs a max and min consensus operation. If the stopping condition holds, it scales the solution according to the quantization level.

% This means that each node is able to calculate the quantized average of each node’s estimate in finite time by processing and transmitting quantized messages, with precision determined by the quantization level. 

% \todo{use flowchart } 

% \todo{
% $ \frac{1}{n}\sum_{i=1}^n  \nabla f_j(x_j[k+1]) = \Delta\frac{1}{n}\sum_{i=1}^n \Bigl \lfloor \frac{\nabla f_j(x_j[k+1])}{\Delta} \Bigr \rfloor+ \epsilon, \epsilon \le \Delta  $}

% \todo{
% $-\Delta \le q^s_j[\lambda+1] -\Delta\frac{1}{n}\sum_{i=1}^n \Bigl \lfloor \frac{\nabla f_j(x_j[k+1])}{\Delta} \Bigr \rfloor \le \Delta$}

%   %$x_j[k+1] =  \lfloor  \frac{x_j[k+\frac{1}{3}]}{\Delta} \rfloor \Delta$; $x_j[k+\frac{1}{3}]$

% \todo{
% $ \Delta = 0.01,   \lfloor \frac{0.013}{\Delta}  \rfloor = 1, 1*\Delta= 0.01$}

% \todo{
% $ 0.013-0.01 = 0.003 \le  \Delta * 0.5 = 0.005$ }

% \todo{
% $ \Delta = 0.01,   \lfloor \frac{0.019}{\Delta}  \rfloor = 1, 1*\Delta= 0.01,$}

% \todo{
% $ 0.019-0.01 = 0.009 \le  \Delta * 1 = 0.01$ }

% \begin{remark}\label{operation_remark}
%     \todo{compare with previous works + emphasize on advantages: Mention Wei-Themis work + mention Salapaka work} \\ \cite{2022:Jiang_Charalambous}
%     + 
%     \todo{we can accelerate??? - will be mentioned in an extended -- this is Wei's comment} \textcolor{blue}{Wei: Not sure any more since algorthrim is changed.}
% \end{remark}

\subsection{Convergence of Algorithm~\ref{algorithm_1}}\label{alg_convergence}

We now analyze the convergence  of Algorithm~\ref{algorithm_1}. 
Specifically, we show that during the operation of Algorithm~\ref{algorithm_1}, the variable $x_i^{[k]}$ of each node $v_i \in \mathcal{V}$ converges to a neighborhood of the optimal solution $x^*$ with linear convergence rate. 
%\subsection{Step-Size Upper Bound}
% \textcolor{red}{Wei: neglect the notions I use (they are convenient for me to finish the proof in the first round), I can easily change them later. Also, polish later}
%\end{equation}
In Step 1) of Algorithm~\ref{algorithm_1}, for the convenience of presentation of mathematics in this section, we denote $ z_i^{[k+1]} \coloneqq x_i^{[k+\frac{1}{2}]} $. Thus, the Steps 1) and 2) in Algorithm~\ref{algorithm_1} are changed to:
\begin{align}
z_i^{[k+1]} =&  x_i^{[k]} - \alpha \nabla f_i(x_i^{[k]}), \label{x_medium_update}\\
 x_i^{[k+1]} =& \text{Algorithm}~\ref{algorithm_1a}(z_i^{[k+1]} , D, \Delta ).\label{x_final_update}
\end{align}
%Here, we are interested in developing conditions for the step-size upper bound theoretically. 

%\todo{for exact algorithm} 

From~\eqref{x_final_update} and the property of Algorithm~\ref{algorithm_1a}, we have
\begin{equation}\label{algorithm1a_ac}
x_i^{[k+1]} = \frac{1}{n}\sum_{i=1}^{n}\Delta \Bigl \lfloor \frac{z_i^{[k+1]} }{\Delta} \Bigr \rfloor + \varrho_i^{[k+1]}, \ \text{where} \ \| \varrho_i^{[k+1]}\| \le \Delta, 
\end{equation}
for every $k \ge 0$, where $\varrho_i^{[k+1]}$ is the total error due to Algorithm~\ref{algorithm_1a} calculating the quantized average of the node's initial quantized states at optimization step $k$.
%Denote $ \hat{x}^{k+1} \coloneqq \frac{1}{n}\sum_{i=1}^{n} x_i^{k+1}, k\ge 0 $. 

In addition,
from the quantization definition, we have 
%\begin{equation}
\begin{align}
%\nabla f_j(x_j^{k+1}) =& \Delta \Bigl \lfloor \frac{\nabla f_j(x_j^{k+1})}{\Delta} \Bigr \rfloor  + \varrho_j^k, 0\le \varrho_j^k \le \Delta, \forall i, \label{eq_varepsilon}\\
  x_i^{[k]} =&  \Delta \Bigl \lfloor \frac{x_i^{[k]}}{\Delta} \Bigr \rfloor + \epsilon_i^{[k]}, \ \text{where} \ 0 \le \epsilon_i^{[k]} \le \Delta, \label{eq_epsilon}
\end{align}
for every $v_i \in \mathcal{V}$, where $\epsilon_i^{[k]}$ is the error due to applying asymmetric quantization to the value $x_i^{[k]}$ (see Section~\ref{sec:preliminaries}) at time step $k$.
Denote $$\hat{z}^{[k+1]}  \coloneqq \frac{1}{n}\sum_{i=1}^{n} z_i^{[k+1]}, \hat{x}^{[k+1]} \coloneqq \frac{1}{n}\sum_{i=1}^{n} x_i^{[k+1]}, k\ge 0.$$ Based on the quantizer property~\eqref{eq_epsilon}, it is easy to have 
\begin{align}
 \hat{z}^{[k+1]} - x_i^{[k+1]}   =& \frac{1}{n}\sum_{i=1}^{n} z_i^{[k+1]} - \frac{1}{n}\sum_{i=1}^{n}\Delta \Bigl \lfloor \frac{z_i^{[k+1]} }{\Delta} \Bigr \rfloor -\varrho_i^{[k+1]}\nonumber \\=& \frac{1}{n}\sum_{i=1}^{n}\epsilon_i^{[k+1]}-\varrho_i^{[k+1]} \le 2 \Delta.\label{x_hatz}
\end{align}
%\begin{equation*}
%\hat{x}^{k+1} \coloneqq \frac{1}{n}\sum_{i=1}^{n} x_i^{k+1}, k\ge 0.
%\end{equation*}

%\begin{equation}
%F(\hat{x}^{k+1}) \le F(\hat{x}^{k}) +  \langle \hat u^k, \hat{x}^{k+1}-\hat{x}^{k} \rangle + \frac{L}{2}\|\hat{x}^{k+1}-\hat{x}^{k} \|^2,
%\end{equation}
%which also means 
%\begin{equation}
%\tilde{F}^{k+1} \le \tilde{F}^{k} +  \langle \hat u^k, \hat{x}^{k+1}-\hat{x}^{k} \rangle + \frac{L}{2}\|\tilde{F}^{k+1}-\tilde{F}^{k} \|^2,
%\end{equation}
%From~\eqref{hatx_calcu2}, we have 
%\begin{align}
%\tilde{F}^{k+1} \le& \tilde{F}^{k} -\frac{\alpha}{n} \langle \hat u^k, u^k \rangle +\langle \hat u^k, \phi^k \rangle  + \frac{L}{2}\|\frac{\alpha}{n}u^k   -\phi^k \|^2 \label{hatx_calcu3} \\ 
%\le& \tilde{F}^{k} -\alpha \langle \hat u^k, u^k \rangle +\langle \hat u^k, \phi^k \rangle  + \frac{\alpha^2L}{2} \|u^k\|^2  + \frac{L}{2} \|\phi^k\|^2  - \alphaL  \langle  u^k, \phi^k \rangle \nonumber 
%\end{align}
%
%We also have 
%\begin{align}
%-\alpha \langle \hat u^k, u^k \rangle = & -\alpha \langle  u^k + \hat u^k - u^k, u^k \rangle \nonumber \\
%= & -\alpha\|u^k\|^2 + \alpha \langle  u^k -\hat u^k , u^k \rangle.
%\end{align}

%\eqref{hatx_calcu3} changes to
%\begin{align}
%\tilde{F}^{k+1} 
%\le& \tilde{F}^{k} -\alpha (1-\frac{\alphaL}{2}) \|u^k\|^2 +\alpha \langle  u^k -\hat u^k , u^k \rangle +\langle \hat u^k, \phi^k \rangle \nonumber\\&  + \frac{L}{2} \|\phi^k\|^2  - \alphaL  \langle  u^k, \phi^k \rangle \nonumber 
%\end{align}

\begin{lemma}\label{lemma_uk_phik}
	For $ k \ge 1 $, the following inequalities hold:
	\begin{align}
	\| \hat x^{[k]}-\hat{z}^{[k]}  \|\le & 2\Delta, \label{uk_hatuk}\\
	\| x_i^{[k]}- \hat x^{[k]} \| \le& 4\Delta. \label{phik}
	\end{align}
\end{lemma}
\begin{pf}
%	See  Appendix~\ref{appen_lemma}.
%  Then,
From~\eqref{algorithm1a_ac}, we have 
%\vspace{-0.1cm}
%\begin{equation}
\begin{align}
\| \hat x^{[k]}-\hat{z}^{[k]}  \| =& \|\frac{1}{n}\sum_{i=1}^{n} (x_i^{[k]}-z_i^{[k]}) \| \nonumber\\ =&\|\frac{1}{n}\sum_{i=1}^{n}( \varrho_i^{[k]} - z_i^{[k]}+ \frac{1}{n}\sum_{i=1}^{n} \Delta \Bigl \lfloor \frac{z_i^{[k]} }{\Delta} \Bigr \rfloor)  \|\nonumber\\
\le&\|\frac{1}{n}\sum_{i=1}^{n} \varrho_i^{[k]}  \| + \|\frac{1}{n}\sum_{i=1}^{n}(  \Delta \Bigl \lfloor \frac{z_i^{[k]} }{\Delta} \Bigr \rfloor -z_i^{[k]})  \| \nonumber\\ \le& 2\Delta,
\end{align}
%\end{equation}
%\begin{equation}
%\| \nabla f_i(x^k_i) - \nabla f_i(\hat x^k)\|\le  L_i\| x^k_i - \hat x^k\| \le  L_i\varepsilon_k,
%\end{equation}
which proves~\eqref{uk_hatuk}.

Based on~\eqref{x_hatz} and \eqref{uk_hatuk}, we have 
\begin{equation*}
\| x_i^{[k]}- \hat x^{[k]} \| \le \| x_i^{[k]}- \hat{z}^{[k]} \| + \| \hat{z}^{[k]}- \hat x^{[k]} \| \le 4\Delta,
\end{equation*}
which can prove~\eqref{phik}. The proof is finished.
\end{pf}

%Denote $ u^k = \sum_{i=1}^{n} \nabla f_i(x_i^{k}), \hat u^k = \sum_{i=1}^{n} \nabla f_i(\hat{x}_i^{k}) $. 
Denote $$ u^{[k]} := \sum_{i=1}^{n} \nabla f_i(x_i^{[k]}), \hat u^{[k]} := \sum_{i=1}^{n} \nabla f_i(\hat x^{[k]}). $$
%From Assumption~\ref{assup_convex} we know the  objective function $ F(x) $ in~\eqref{problem_initial} is $ L $-smooth.
From Assumption~\ref{lipsch_str_conv} with ~\eqref{phik}, we have
\begin{align}
&\| \nabla f_i(x_i^{[k]}) - \nabla f_i(\hat x^{[k]}) \| \le  L_i \| x_i^{[k]} -\hat  x^{[k]}\|\le  4L_i\Delta \\
&\|u^{[k]} - \hat u^{[k]} \| \le  \sum_{i=1}^{n}\| \nabla f_i(x_i^{[k]})-\nabla f_i(\hat{x}^{[k]}) \| \nonumber \\ &\le  \sum_{i=1}^{n} L_i \| x_i^{[k]} -\hat  x^{[k]}\| \le  \AR{4L\Delta, L = \sum_{i=1}^{n} L_i. \label{uk_hat}}
\end{align}

Denote $ e^{[k]} = \hat{x}^{[k]}-\hat{z}^{[k]}, \hat \alpha := \frac{\alpha}{n} $. 
From~\eqref{uk_hatuk} we have $ \|e^{[k]}\|\le 2\Delta $. 
% \todo{From \eqref{x_medium_update} we get
% \begin{align}
% \hat{x}^{[k+1]}  =& \hat{z}^{[k+1]} + e^{[k+1]}\nonumber\\
% =& \frac{1}{n}\sum_{i=1}^{n} (x_i^{[k]} - \alpha \nabla f_i(x_i^{[k]})) + e^{[k+1]}\nonumber\\
% =&\hat{x}^{[k]} - \hat \alpha u^{[k]} + e^{[k+1]},  k\ge 1. \label{hatx_calcu2}
% \end{align}}
From \eqref{x_medium_update} we get $\hat{x}^{[k+1]}  = \hat{z}^{[k+1]} + e^{[k+1]} = \frac{1}{n}\sum_{i=1}^{n} (x_i^{[k]} - \alpha \nabla f_i(x_i^{[k]})) + e^{[k+1]} = \hat{x}^{[k]} - \hat \alpha u^{[k]} + e^{[k+1]}, \ k\ge 1.$
%From~\eqref{algorithm2_x_update},
%based on the property of Algorithm~\ref{algorithm_fterc_termination}, inside each optimization iteration $ k $, denote $ \epsilon^k \coloneqq \frac{1}{n}\sum_{j=1}^{n} \epsilon_j^k, 0\le \epsilon^k \le \Delta $,  we have in finite-time that
%\begin{equation}\label{x_update}
%z_i^{k+1} = \frac{1}{n}\sum_{j=1}^{n} \Delta \Bigl \lfloor \frac{\nabla f_j(x_j^{k+1})}{\Delta} \Bigr \rfloor  = \frac{1}{n}\sum_{j=1}^{n} \nabla f_j(x_j^{k+1}) - \epsilon^k, k\ge 0,
%\end{equation} 
%which also means $ z_i^{k}  = z_j^{k}, k\ge 1 $. Therefore, we have
%\begin{align*}
%\frac{1}{n} \sum_{i=1}^{n} ( z_i^{k} -\nabla f_i(x_i^k)  ) = z_i^{k} - \frac{1}{n} \sum_{i=1}^{n}\nabla f_i(x_i^k) =- \epsilon^k, k\ge 1.
%\end{align*}

%As a consequence, \eqref{hatx_calcu} changes to
%\begin{align}\label{hatx_calcu2}
%\hat{x}^{k+1} =\hat{x}^{k} - \alpha u^k +\phi_k - n\alpha\epsilon^k, k\ge 1.
%%\hat{x}^{k+1} =\hat{x}^{k} - \frac{\alpha}{n} \sum_{i=1}^{n} \nabla f_i(x_i^k) +
%%\phi_k, k\ge 1.
%\end{align}

At present, we are ready to provide the result of the step-size upper bound and algorithm convergence rate in the following theorem. 
However, due to space limitations we omit the proof of Theorem~\ref{theorem_convergence_stronglyConvex}. 
It will be available at an extended version of our paper. 

\begin{theorem}\label{theorem_convergence_stronglyConvex} 
	Under Assumptions~\ref{str_conn}--\ref{Diam_known}, when the step-size $ \alpha $ satisfies $ \alpha \in (\frac{n(\mu + L)}{4\mu L}, \frac{2n}{\mu + L}) $ and $ \delta \in (0,\frac{n[4 \alpha \mu L -n(\mu + L) ]}{2 \alpha [n(\mu + L)-2  \alpha \mu L]} ) $ where  \AR{$ L =  \sum_{i=1}^{n}L_i ,  \mu = \sum_{i=1}^{n}\mu_i$},
    Algorithm~\ref{algorithm_1} generates a sequence of points $ \{x^{[k]}\}$ (i.e., the variable $x_i^{[k]}$ of each node $v_i \in \mathcal{V}$) which satisfy
	\begin{align}\label{linear_convergence}
	\| \hat{x}^{[k+1]} - x^{*}\|^2 
	<
	\vartheta\| \hat{x}^{[k]} - x^{*}  \|^2   + \mathcal{O}(\Delta^2) ,
	\end{align}
	where $ \Delta $ is the quantizer and
%	$ \vartheta \coloneqq 2(1+\frac{\alpha\delta}{n} )(1- \frac{2\alpha\sigma_2}{n} ) \in (0,1), \mathbb{O}({\varepsilon_k^{2}}) =2 (1+\frac{\alpha^2L^2}{n^2}  +\frac{2\alpha L^2}{n\delta}) {\varepsilon_k^{2}}  $ and $ \varepsilon^k \rightarrow 0, k \rightarrow \infty$.
	\begin{subequations}\label{throrem1}
		\begin{align}
		\vartheta := & 2(1+\frac{\alpha\delta}{n} )(1- \frac{2\alpha\mu L}{n(\mu + L)} ) \in (0,1), \label{throrem1_a}\\
		 \mathcal{O}(\Delta^2) 
		 =&(8
		 +\AR{32 \hat{\alpha}^2 L^2 + \frac{32 \hat{\alpha} L^2}{\delta}}) \Delta^2. \label{throrem1_b} 
		\end{align}
	\end{subequations}
\end{theorem}
% \begin{pf}
% See Appendix~\ref{appen_theorem}.
% \end{pf} 
Theorem~\ref{theorem_convergence_stronglyConvex} shows that Algorithm~\ref{algorithm_1} converges linearly to a neighborhood of the optimal solution. 
This neighborhood is determined by the quantization level $\Delta$.
\begin{remark}
% \todo{When $  \alpha \in (\frac{n(\mu + L)}{4\mu L}, \frac{2n}{\mu + L})  $, based on the proof in Appendix~\ref{appen_theorem}, we always have $ \frac{n[4 \alpha \mu L -n(\mu + L) ]}{2 \alpha [n(\mu + L)-2  \alpha \mu L]} > 0  $; in other words, $  \delta $ exists.}
Based on the Theorem~\ref{theorem_convergence_stronglyConvex}, if we have that $ \alpha \in (\frac{n(\mu + L)}{4\mu L}, \frac{2n}{\mu + L}) $, then we always have $ \frac{n[4 \alpha \mu L -n(\mu + L) ]}{2 \alpha [n(\mu + L)-2  \alpha \mu L]} > 0  $, thus, $  \delta $ exists.
Also, we always have $ \vartheta \in (0,1) $.
For the step-size interval $ \alpha \in (\frac{n(\mu + L)}{4\mu L}, \frac{2n}{\mu + L}) $ to be  always not empty, we need $ \frac{n(\mu + L)}{4\mu L} < \frac{2n}{\mu + L} $. 
This means that we need $ (L-\mu)^2 < 4\mu L $. 
Herein, we could provide a sufficient condition to have this interval be always available. 
Due to $ \mu \le L $, if we get $ (L-\mu)^2 < 4\mu ^2 $, then, we always have $ (L-\mu)^2 < 4\mu L $.
This means that when $ L < 3\mu $, the interval for the step-size $ \alpha $ is always not empty. 
\end{remark}
% ===============================================
%
%
% SIMULATIONS
%
%
% ===============================================
\section{Simulation Results}\label{sec:results}

% \todo{MENTION PARAMETERS DIAMETER ALPHA ETC -- WITH WEI}

In this section, we present simulation results in order to demonstrate the operation of Algorithm~\ref{algorithm_1}. 
We focus on a random digraph of $20$ nodes and show how the nodes’ states converge to the optimal solution for various quantization levels. 
Furthermore, we present comparisons against existing algorithms, and we emphasize on the improvements introduced by Algorithm~\ref{algorithm_1}. 

\begin{figure}[t]
    \centering
    \includegraphics[width=\linewidth]{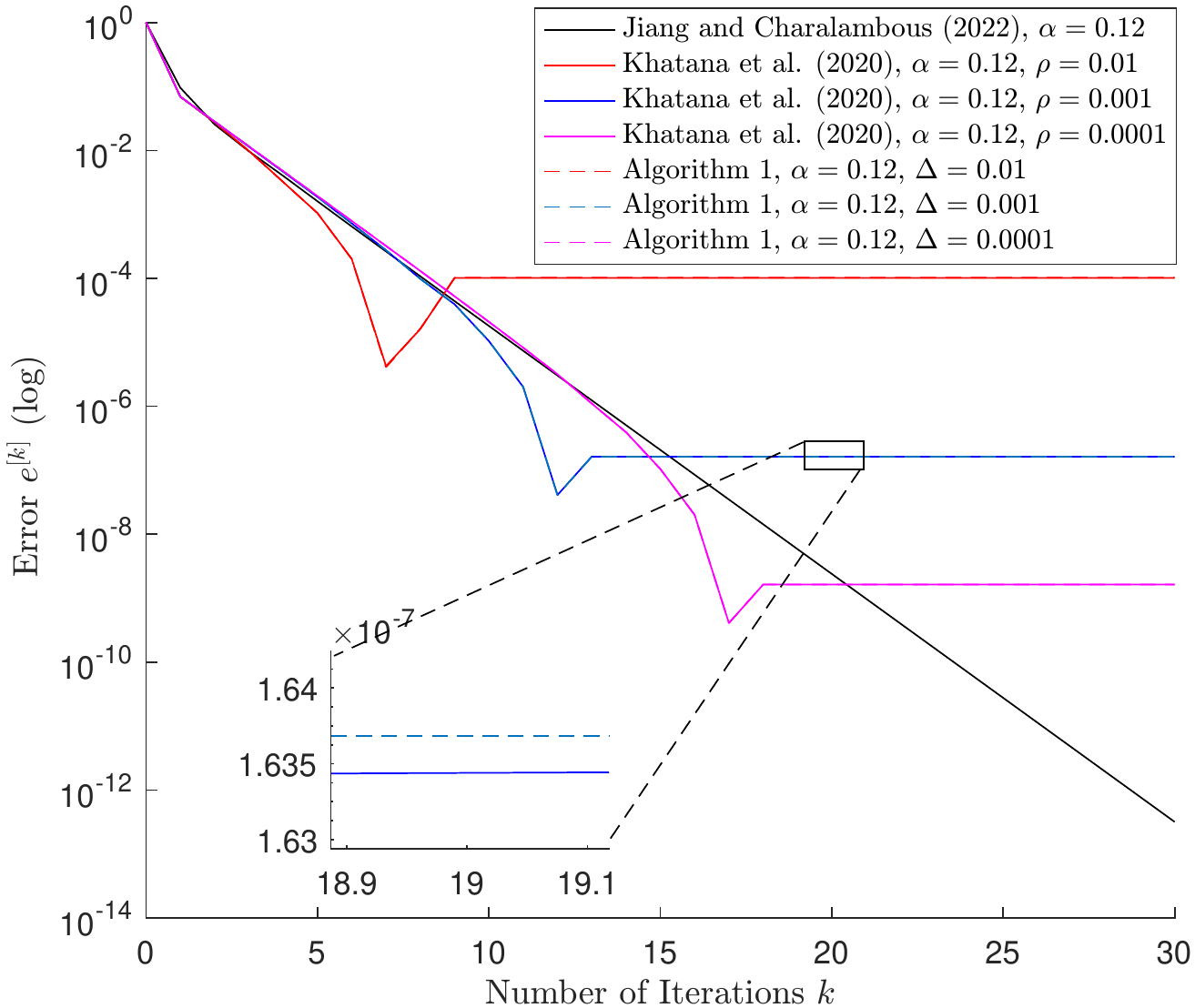}
    \caption{Comparison of Algorithm~\ref{algorithm_1} for different quantization levels with \cite{2022:Jiang_Charalambous, Khatana:2020ACC}.} 
    \label{20_nodes_error}
\end{figure}

In Fig.~\ref{20_nodes_error}, we compare the operation of Algorithm~\ref{algorithm_1} for different quantization levels with \cite{2022:Jiang_Charalambous, Khatana:2020ACC}. 
We plot the error $e^{[k]}$ in a logarithmic scale against the number of iterations. 
The error $e^{[k]}$ is defined as 
\begin{equation}\label{eq:distance_optimal}
    e^{[k]} = \sqrt{ \sum_{j=1}^n \frac{(x_j^{[k]} - x^*)^2}{(x_j^{[0]} - x^*)^2} } , 
\end{equation}
where $x^*$ is the optimal solution of the optimization problem $P1$. 

We can see that Algorithm~\ref{algorithm_1} exhibits almost equal performance compared with \cite{Khatana:2020ACC}, for the case where the quantization level is equal to the pre-specified tolerance value $\rho$ (see \cite{Khatana:2020ACC}). 
However, our proposed algorithm uses quantized values and hence can be used in channels with limited/finite capacity. 
% and no error \todo{why no error?} is introduced for having finite-time consensus (which is the main reason of the existence of the pre-specified tolerance value).
Furthermore, Algorithm~\ref{algorithm_1} exhibits comparable performance compared with \cite{2022:Jiang_Charalambous}, even though the results of \cite{2022:Jiang_Charalambous} do not have an error floor. 
Note, however, that Algorithm~\ref{algorithm_1} is able to approximate 
% in finite time \todo{Alg. 1a is finite time, Alg 1 convergence is asymptotic, not finite-time} 
the optimal solution with precision that depends on the quantization level. 
This means that if we reduce the value of the quantization level, nodes are able to approximate the optimal solution with higher precision. 
%However, note here that Algorithm~\ref{algorithm_1} utilizes messages with much lower precision compared with \cite{2022:Jiang_Charalambous, Khatana:2020ACC}. 
Specifically, for the method in \cite{2022:Jiang_Charalambous}, forming the Hankel matrix and performing additional computations when the matrix loses rank, requires the exact values from each node. 
This translates to nodes exchanging messages of infinite capacity. 
Thus, the main advantage of Algorithm~\ref{algorithm_1} compared to \cite{2022:Jiang_Charalambous} is that nodes operate with quantized values (while in \cite{2022:Jiang_Charalambous} nodes exchange values of infinite prevision). 

% Note here that the main purpose of this paper is to introduce a novel distributed otiimaztion algorithm ... 
% An extensive analysis of ... will be presented in an extended version of this paper. 

%As a result, Algorithm~\ref{algorithm_1} is able to overcome this issue by allowing nodes exchanging messages with much lower precision while maintaining comparable performance compared with \cite{2022:Jiang_Charalambous, Khatana:2020ACC}. 

% \todo{how much lower commuication we need compared to Themis}

% ===============================================
%
%
% CONCLUSIONS
%
%
% ===============================================
\section{Conclusions and Future Directions}\label{sec:conclusions}

In this paper, we focused on designing a communication-efficient algorithm for the unconstrained distributed optimization problem. 
Specifically, each node performs a gradient descent step, and then performs a finite-time calculation of the quantized average of every node’s estimate in the network. 
% Therefore, in essence, the algorithm \emph{approximately} mimics \todo{why mimic, any proof?} - \AR{MENTIONED IN KHATANA} the centralized gradient decent algorithm. 
This algorithm, to the best of the authors' knowledge, is the first distributed optimization algorithm which relies on a finite time quantized coordination mechanism which operates over directed graphs, and the only one with quantization used in this context of approximating the centralized gradient decent algorithm. 

\bibliography{ifacconf}   

\end{document}